\documentclass[aps,prl,twocolumn,groupedaddress]{revtex4-1}
\usepackage{epsfig}
\usepackage{subfigure}
\usepackage{graphicx}
\usepackage{amsfonts}
\usepackage[figuresright]{rotating}
\usepackage{amssymb}
\usepackage{amsmath}
\usepackage{psfrag}
\usepackage{esint} 
\usepackage{bm}
\usepackage{comment}
\usepackage[colorlinks=true,linkcolor=blue,anchorcolor=blue,citecolor=blue,urlcolor=blue]{hyperref}

\def\be{\begin{equation}} \def\ee{\end{equation}}
\def\bea{\begin{eqnarray}} \def\eea{\end{eqnarray}}

\def\nn{\nonumber}

\def\k{{\bf k}}
\def\p{{\bf p}}
\def\K{{\bf K}}
\def\Q{{\bf Q}}
\def\A{{\bf A}}
\def\r{{\bf r}}
\def\g{\tilde{\gamma}}

\def\bnabla{{\boldsymbol \nabla}}

\newcommand{\ehat}{\mathbf{\hat{e}}}

\makeatletter
\newcommand*{\balancecolsandclearpage}{%
  \close@column@grid
  \clearpage
}
\makeatother

\begin{document}
\title{Monopole Charge Density Wave States in Weyl Semimetals} 
\author{Eric Bobrow}
\affiliation{Department of Physics and Astronomy, The Johns
Hopkins University, Baltimore, Maryland 21218, USA}
\author{Canon Sun}
\affiliation{Department of Physics and Astronomy, The Johns
Hopkins University, Baltimore, Maryland 21218, USA}
\author{Yi Li}
\affiliation{Department of Physics and Astronomy, The Johns
Hopkins University, Baltimore, Maryland 21218, USA}
\date{September 19, 2019}

\begin{abstract}
We study a new class of topological charge density wave states exhibiting monopole harmonic symmetries.
The density-wave ordering is equivalent to
pairing in the particle-hole channel due to Fermi surface nesting under interactions.
When electron and hole Fermi surfaces carry different Chern numbers, the particle-hole pairing exhibits a non-trivial Berry phase inherited from band structure topology independent of concrete density-wave ordering mechanism.
The associated density-wave gap functions become nodal, and the net nodal vorticity is determined by
the monopole charge of the pairing Berry phase.
The gap function nodes become zero-energy Weyl nodes of the bulk spectra of quasi-particle excitations.
These states can occur in doped Weyl semimetals with nested electron and hole Fermi surfaces enclosing
Weyl nodes of the same chirality in the weak coupling regime.
Topologically non-trivial low-energy Fermi arc
surface states appear in the density-wave ordering
state as a consequence of the emergent zero-energy Weyl nodes.
\end{abstract}
\maketitle

{\it Introduction.} --
Charge density wave (CDW) ordering, the spontaneous ordering of electron density or bond strength, is an important phenomenon in correlated electron systems \cite{gruner1988,heeger1988}. The broken translational symmetry of CDW ordering often arises from a Peierls instability, which is driven by electron-phonon interactions between nested Fermi surfaces that lead to the softening of phonon modes and accompanying periodic lattice distortions
\cite{thorne1996}.
Novel topological electron excitations can exist at defects in the CDW
order, such as half-fermion modes localized around the domain walls
of the Peierls distortion in the one-dimensional polyacetylene
chain \cite{heeger1988,su1980}.
The CDW instability may also be driven by electron-electron interactions,
as studied in the context of high-T$_c$ cuprates \cite{nayak2000,fujita2014,fradkin2015}.
Analogously to unconventional superconductivity, CDW order may also
possess unconventional symmetries, forming a non-trivial representation
of the lattice symmetry group.
For example, CDW order with a $d$-wave form factor was proposed
to compete and coexist with superconductivity
\cite{nayak2000,fujita2014}.

Study of the Berry phase of Bloch wave states in lattice systems has led
to the discovery of a plethora of topological states, such as quantum
anomalous Hall insulators \cite{Haldane1988,Chang2013} and topological
insulators \cite{Kane2005,Kane2005a,Bernevig2006a,Bernevig2006}.
Furthermore, the discovery of Weyl semimetals \cite{Murakami2007,Wan2011,Burkov2011,Yang2011,Xu2011,Meng2012,
Witczak-Krempa2012,Cho2012,Hosur2012,Fang2012,Halasz2012,Son2013,Hosur2013,wang2013a,Vazifeh2013,Nandkishore2014,Hosur2014,Potter2014,Haldane2014,Wei2014,Yang2014,Weng2015,Burkov2015,Borisenko2015,Xiong2015a,Xu2015,Xu2015a,Huang2015,Lv2015,Lu2015,
Xu2016,Yan2017,Armitage2018,Li2018} has opened up a new avenue for studying
topological phases in semi-metallic systems. As in quantum anomalous Hall insulators where
a Chern number structure arises from quantized Berry flux over
the two-dimensional Brillouin zone, in three-dimensional semimetals the Fermi surfaces have a Chern number structure due to the Weyl points acting as sources or sinks of Berry flux.

After doping, magnetic Weyl semimetals can host monopole harmonic superconductivity, a novel class of topological states.
As opposed to typical unconventional superconductors,
such as $d$-wave high $T_c$ cuprates, and $p$-wave superfluid $^3$He,
in monopole harmonic superconductors the gap function $\Delta(k)$ cannot be described by spherical harmonics and their lattice counterparts \cite{Li2018}.
Instead, these systems carry ``pairing monopole charge", a generalization
of Berry phase from single-particle states to a two-particle order parameter.
When the pairing occurs between two Fermi surfaces with opposite Chern
numbers, which can be the case when the enclosed Weyl points have opposite chiralities,
Cooper pairs acquire non-trivial Berry phase structure.
As a result, the gap function cannot be well defined over the entire Fermi surface.
Consequently, the Fermi surface becomes nodal with total vorticity determined by
the pairing monopole charge associated with the two-particle
Berry phase.

In this article, we study non-trivial Berry phase structure for a class of order parameter gap functions lying in the particle-hole channel.
As an example that can be realized in a doped Weyl semimetal, CDW ordering will be considered.
When two nested Fermi surfaces, one electron-like and one hole-like,
carry different Chern numbers, the CDW order formed between
these two Fermi surfaces inherits non-trivial band structure topology that can be seen in its gap function $\rho(\bf{k})$.
As with the gap function of monopole harmonic superconductivity, $\rho(\bf{k})$ cannot
be globally well defined in momentum space and becomes nodal.
The nontrivial Berry flux enforces a nonzero total vorticity of
$\rho(\bf{k})$ determined by the difference in Chern number
between the two nested Fermi surfaces that is independent of the
concrete mechanism for CDW ordering.
The nodes of the CDW gap function emerge as new Weyl nodes in the
low-energy quasi-particle spectra that are distinct from the original
band structure Weyl points.
The chiralities of the emergent quasi-particle Weyl nodes are determined
by the band structure in which the single-particle Weyl points have been shifted away from
the Fermi surfaces after doping.

{\it Gap function Berry flux and nodes for CDW ordering.} --
We begin with a minimal description of a pair of electron-like and hole-like Fermi surfaces, which carry opposite Chern numbers and are well nested.
Such Fermi surfaces can be realized in a 3D Weyl semimetal system around two Weyl points
of the same chirality.
Consider two Weyl points of
positive chirality located at $\K^+_{e}$ and $\K^+_h$
with energies $-E_0$ and $E_0$ respectively and Fermi energy at $\mu=0$.
A hole-like Fermi surface denoted $\text{FS}_{h,C}$ is centered
around $\K^+_{h}$ with 
Chern number $C$ or, equivalently, monopole charge $q=\frac{1}{2} C$.
Similarly, an electron-like Fermi surface denoted $\text{FS}_{e,-C}$ is centered around $\K^+_e$ with Chern number $-C$.
An example of a system with this Fermi surface structure is considered in Eq. \eqref{eq:Hamiltonian}
below.


Since $\text{FS}_{h,C}$ and $\text{FS}_{e,-C}$ are well nested, they favor
a CDW instability, inter-Fermi surface particle-hole
pairing, under repulsive interactions.
The two-particle CDW order parameter exhibits a non-trivial Berry flux
quantization, which can be seen as follows.
After projecting to the low-energy Fermi surfaces, the electron creation
operators on $\text{FS}_{h,C}$ and $\text{FS}_{e,-C}$ can be defined as
$
\alpha^{\dagger}_{\pm}(\p) =\sum_{a=\uparrow,\downarrow}\xi_{\pm,a}(\p)c^{\dagger}_a(\K_{h(e)}^+ +\p)
$,
where $\p$ is the momentum relative to the Weyl node at $\K^+_{h(e)}$,
$a$ refers to the spin or pseudospin degrees of freedom,
and $\xi_{\pm}(\p)$ is the spinor eigenfunction carrying monopole
charge $\pm q$.
Here $\p$ lies on the surface $S$ that results from shifting
$\text{FS}_{e,-C}$ by $-\K_{e}^+$ towards the origin.
We define the particle-hole channel pairing operator
\bea
P_{-+}(
\p)=\alpha_{-}^{\dagger}(\p)\alpha_{+}(\p),
\eea
which creates an electron on $\text{FS}_{e,-C}$ and a hole on
$\text{FS}_{h,C}$.
The single-particle Berry connection is $\A_{\pm}(\p)=\sum_a
i\xi^{*}_{\pm,a}(\p)\bnabla_{\p} \xi_{\pm,a}(\p)$, and the
Berry flux penetrating $\text{FS}_{\pm}$
is given by $\oiint_{\text{S}} d\p \cdot \bnabla_{\p}
\times \A_{\pm}(\p)=\pm 4 \pi q$.
It can be shown that the pairing Berry connection associated
with $P_{-+}(\p)$ is $\A_{-+}(\p)=\A_{-}(\p)-\A_{+}(\p)$,
and the net pairing Berry flux through $S$ is
$
\oiint_{S} d\p
\cdot \bnabla_{\p}\times\A_{-+}(\p)=4\pi q_{CDW}
$,
where $q_{CDW}=-2q$.

The non-zero Berry flux through $S$ leads to a non-trivial vortex
structure for the CDW gap function.
The CDW interaction Hamiltonian after mean-field decomposition
is expressed as
\bea
H_{\rho}= \sum_\p \rho_{-+} (\p) P_{-+}(\p) +\rho_{+-}(\p) P_{+-} (\p),
\eea
where $P_{+-}=P_{-+}^\dagger$.
The gap function $\rho_{-+}$ is conjugate to the CDW operator
$P_{-+}(\p)$, and $\rho_{+-}=\rho_{-+}^*$.
Because of the non-trivial gauge
field $\A_{-+}$, $\rho_{-+}(\p)$  cannot be globally well defined on $S$.
This follows from examining the gauge invariant circulation field
$\mathbf{v}_{-+}(\p)=\nabla \phi_{-+}(\p)-\A_{-+}(\p)$,
where $\phi_{-+}$ is the phase of $\rho_{-+}$.
$\mathbf{v}_{-+}$ is well defined except at the nodes of
$\rho_{-+}(\p)$, and each node has an integer-valued
vorticity $g_i=\frac{1}{2\pi} \oint_{C_i} d\p \cdot \mathbf{v}_{-+}(\p)$.
$C_i$ is an infinitesimal loop around the node $\p_i$, with
the positive loop direction defined with respect to the local normal vector.
The total vorticity of $\rho_{-+}$ over $S$ is
$\sum_i g_i=2q_{CDW}$,
where the sum is over all the nodes on $S$.
As a consequence, the enclosure of a non-zero net monopole charge
gives rise to nodes of $\rho_{-+}(\p)$ on $S$.
The non-trivial nodal structure necessitates the use of the
monopole harmonic functions \cite{Wu1976}, as opposed to the
usual spherical harmonics 
to describe the order parameter.

\begin{figure}[tbp]
\subfigure[]{\centering \epsfig{file=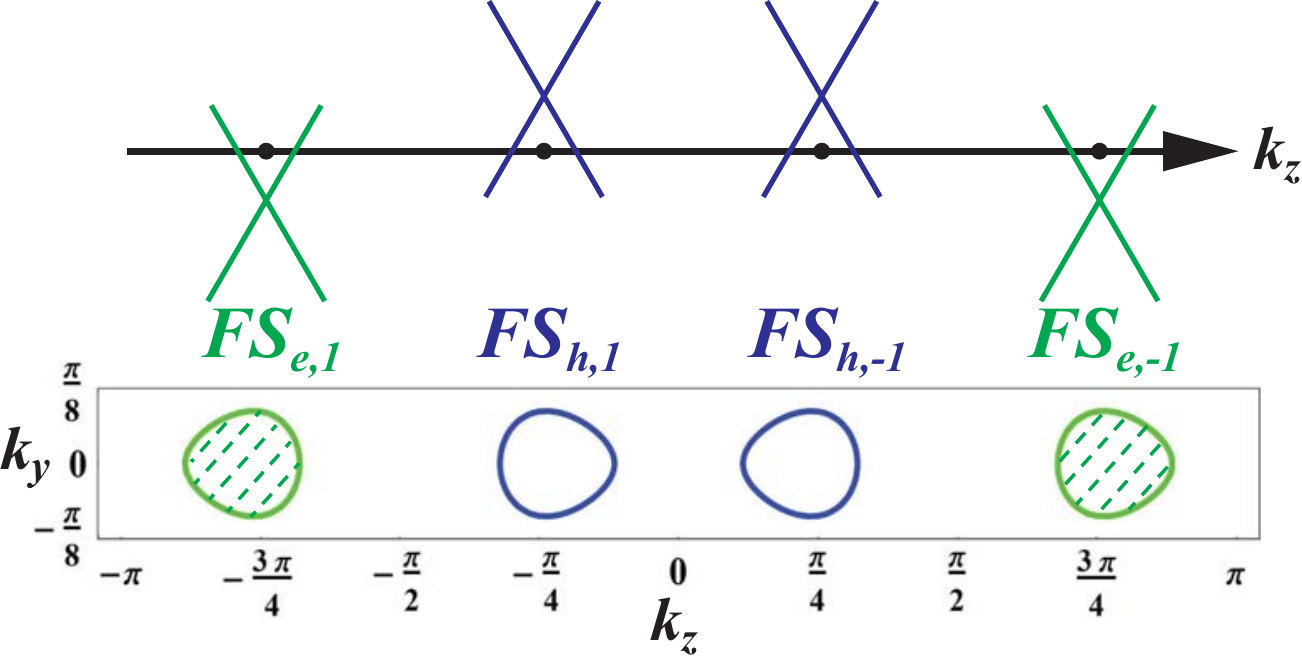,width=0.58\linewidth}}
\subfigure[]{\centering
\epsfig{file=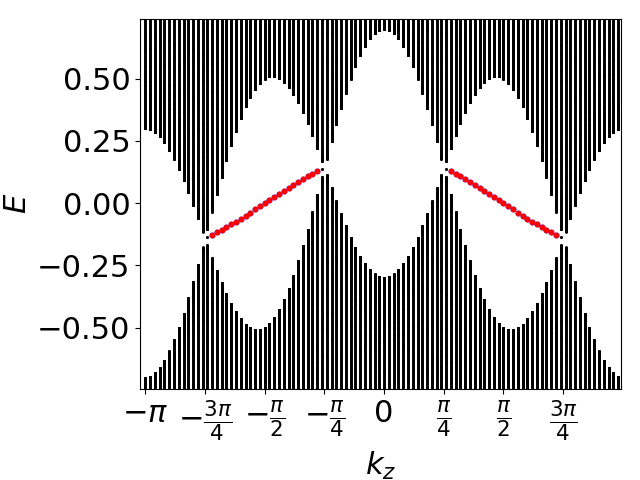,width=0.4\linewidth}}
\caption{
(a) The four Fermi surfaces of the band Hamiltonian in
Eq. (\ref{eq:bloch}) with $V_0/t=0.49$.
They are divided into two well-nested pairs:
FS$_{h,1}$ and FS$_{e,-1}$ which enclose the Weyl points $\K^+_{h,e}$ with
 positive chirality, and FS$_{h,-1}$ and FS$_{e,1}$,
which enclose the Weyl points $\K^-_{e,h}$
with negative chirality.
(b) Energy spectrum as a function of $k_z$ in the absence of the CDW ordering and
$V_0/t=0.2$.
The surface states localized on the $y=0$ boundary are plotted
in red connecting $\K^{-}_e$ and $\K^+_h$, and $\K^-_h$ and $\K^+_e$.
}
\label{fig:cdw_band}
\end{figure}



{\it Doped Weyl Semimetal with Nested Fermi Surfaces.} --
To demonstrate the above topological nodal structure, we employ
the band Hamiltonian
\begin{equation}\label{eq:Hamiltonian}
H = \sum_{\substack{\k, \\ a, b}} c^\dagger_a(\k) [h(\k) + V(\k)]_{ab} c_b(\k) + H_\rho,
\end{equation}
where $a,b$ refers to the pseudospin degree of freedom, typically
realized by $A$ and $B$ sublattices;
$H_\rho$ is the mean-field Hamiltonian for CDW ordering specified below;
and we assume the chemical potential $\mu=0$. 
The matrix kernel $h(\k)$ of the band Hamiltonian is
\begin{equation}
\begin{aligned}
\label{eq:bloch}
h(\k) &= t_z\bigg(2-\cos k_x - \cos k_y -\gamma + \cos ^2k_z\bigg)\tau_z \\
&+ t_x \sin k_x \tau_x + t_y \sin k_y \tau_y,
\end{aligned}
\end{equation}
with Pauli matrices $\tau_{x,y,z}$ defined in the $A, B$ basis and pseudospin-dependent hopping amplitudes $t_{x,y,z}$. Here $\gamma = 1/2$ controls the location of the Weyl points along $\k_z$. For simplicity, we choose $t_{x,y,z}=t$ in this paper.
The corresponding lattice model giving rise to $h(\k)$ is presented in
Supplemental Materials (S.M.) I.
The momentum-dependent potential $V(\k)$ takes the form
\begin{equation}
V(\k) = V_0 \cos k_z I,
\label{eq:v}
\end{equation}
with $I$ the $2\times 2$ identity matrix.
$V_0$ plays a role similar to a chemical potential by controlling the size
of the Fermi surface.

Without loss of generality, we assume $V_0>0$.
This model possesses four Weyl points, all located on the $k_z$ axis, at $\K_e^-=(0,0,-\frac{3\pi}{4})$, $\K_h^+=(0,0,-\frac{\pi}{4})$, $\K_h^-=-\K_h^+$, and $\K_e^+=-\K_e^-$, where the upper indices $\pm$ refer to the chiralities of the Weyl points and the lower indices $e$ and $h$ refer to whether the Fermi surface associated with the Weyl point is electron-like or hole-like. The potential $V(\k)$ shifts the points $\K_h^-$ and $\K_h^+$ up in energy, forming the respective hole-like Fermi surfaces $\text{FS}_{h,1}$ and $\text{FS}_{h,-1}$, as shown in Fig. \ref{fig:cdw_band} ($a$). Similarly, the points $\K_e^+$ and $\K_e^-$ are shifted down in energy, forming the electron pockets $\text{FS}_{e,-1}$ and $\text{FS}_{e,+1}$.
This model is a modification of the models in Refs. \onlinecite{Wang2016, Yang2011} to allow four Weyl points with nesting.
The electron and hole Fermi pockets enclosing the Weyl points with the
same chirality are nested with the commensurate wavevector $\Q=(0,0,\pi)$.
This nesting condition is satisfied so that portions of the Fermi surface separated by $\Q$ have the same shape.
Under an open boundary condition along the $y$-direction and periodic
boundary conditions along $x$ and $z$, the energy spectrum in the absence of the
CDW ordering is plotted in Fig. \ref{fig:cdw_band} ($b$) as a function of $k_z$
along the $k_x = 0$ cut.
The surface Fermi arc states are shown in red.

CDW ordering is imposed through the mean-field Hamiltonian
\begin{equation}
H_\rho = \sum_{\substack{\k, \\ a,b = A,B}}c^\dagger_a(\k + \Q)\rho_{ab}(\k)
c_b(\k) + h.c.,
\label{eq:cdw}
\end{equation}
where we take $\rho(\k) = \rho \tau_z$ and $\rho$ is the magnitude of
the CDW ordering.
$\rho(\k)$ is diagonal in the sublattice $A$ and $B$ basis, which describes two sublattices with different charge densities.
Below we will see that this CDW ordering does not open a full gap
over the Fermi surface but instead becomes nodal with a non-trivial vorticity.

{\it Topological Nodal CDW.} --
We first consider the CDW gap function connecting FS$_{h,1}$
and FS$_{e,-1}$ which enclose the Weyl nodes $\K^+_h$ and $\K^+_e$,
respectively.
For small $V_0/t$ and $\rho/t$, the Fermi surfaces are close to the Weyl nodes and the single-particle states correspond to the helicity eigenstates satisfying $\hat \p\cdot \tau \xi_\pm=\mp \xi_\pm$, where $\xi_\pm$ corresponds to Berry flux monopole charges of $q=\pm \frac{1}{2}$.
Explicitly, $\xi_\pm$ can be represented as
$\xi_+(\hat \p)=(-\sin\frac{\theta_p}{2}e^{-i \phi_p},
\cos \frac{\theta_p}{2})^T$ and $\xi_-(\hat \p)=(\cos \frac{\theta_p}{2},
\sin \frac{\theta_p}{2}e^{i \phi_p})^T$, where
$\theta_p$ and $\phi_p$ are the polar and azimuthal angles
of $\hat \p$ and a gauge convention has been chosen.
The electron creation operator for the eigenstate on the helical
Fermi surface FS$_{h,1}$ is
$\alpha_{+,h}^\dagger(\p) =\sum_a\xi_{+,a}
(\hat \p) c^\dagger_a(\K^+_h+\p)$,
and that for FS$_{e,-1}$ is
$\alpha_{-,e}^\dagger(\p) =\sum_a \xi_{-,a}
(\hat \p) c^\dagger_a(\K^+_e+\p)$.

The gap function $\rho\tau_z$ in Eq. \eqref{eq:cdw} can now be projected onto the helical
Fermi surfaces FS$_{h,1}$ and FS$_{e,-1}$
, where the projected gap function conjugate to $P_{-+}(\p) = \alpha^\dagger_{-,e}(\p)\alpha_{+,h}(\p)$ is $\rho_{-+}(\p) =-\rho \sqrt{8\pi/3}Y_{q=-1,l=1,m=0}(\theta_p,\phi_p)$, in terms of monopole harmonics $Y_{qlm}$ \cite{Wu1976}. For $\p$ near the north pole of $FS_{e, -1}$, where $\theta_p = 0$, the projected gap function is
$\rho_{-+}^N(\p) = -\rho\sin \theta_p e^{-i\phi_p}$. By applying a gauge transformation, the projected gap function near the south pole, where $\theta_p$, can be similarly shown to be $\rho^S_{-+}(\p) = -\rho\sin\theta_p e^{i\phi_p}$. The projected gap function has nodes at the poles, where $\sin \theta_p = 0$.
After taking into account the contribution of the Berry connection $\A_{-+}(\p)$, the circulation field of the gap function is
${\bf v}_{-+}(\p) = -\cot \theta_p \hat{\phi}_p$.
Integrating ${\bf v}$ around infinitesimal loops near $\theta_p = 0$ and $\pi$
reveals a gap function vorticity of $-1$ near both poles,
hence the total vorticity is $-2$ on the Fermi surface
surrounding FS$_{e,-1}$, consistent with $q_{CDW} = -1$.

The CDW gap function nodes are actually low-energy Weyl points
generated by interactions for the mean-field Hamiltonian. 
Around the nested Fermi surfaces FS$_{e,-1}$ and FS$_{h,1}$,
the low-energy two-band Hamiltonian is
\bea
H_{2band}&=&\sum_\p \psi^\dagger(\p) \Big\{ (t|p|- \mu) \sigma_z
-\rho \sin\theta_\p (e^{-i\phi_p} \sigma_+ \nonumber \\
&+& e^{i\phi_p} \sigma_-)
\Big \} \psi(\p),
\label{eq:heff}
\eea
where $\psi(\p)=(\alpha_{-,e}(\p),\alpha_{+,h}(\p))^T$ and $\mu = -V(\K_e^+ + \p)$.
The interaction-induced Weyl node at the north pole, denoted $\K^+_n$, has positive chirality as can be shown by expanding
$H_{2band}$ about the north pole, where the helical basis is regular.
The south pole, denoted $\K^+_s$, is the site of a singularity
in the helical basis and thus needs to be treated more carefully.
Taking into account the $4\pi$-flux from the Dirac string penetrating
the south pole, or equivalently changing the gauge choice to place the singularity
at the North pole, this Weyl node can also be shown to possess positive chirality as well.

The positive chiralities of $\K^+_{n,s}$ are
in fact determined by the chiralities of the original
band structure Weyl nodes $\K^+_{e,h}$, which are away from the
chemical potential and hence lie in the high energy sector.
Nevertheless, they still determine the chirality of the low-energy Weyl nodes,
independent of the details of the mechanism of the CDW ordering.
Typically, the low-energy physics is not sensitive to the details
at high energy, but the topological structure at low-energy in our case
is indeed inherited from the topology at high energy, and thus
the emergence of the low-energy Weyl fermions are topologically
protected.

Similar analysis can also be performed in parallel for the CDW
ordering connecting the nested Fermi surfaces FS$_{h,-1}$ and
FS$_{e,1}$ surrounding $\K^-_{h,e}$, respectively.
The two low-energy Weyl nodes denoted $\K^-_{n,s}$ on the nested FS$_{h,-1}$ and FS$_{e,1}$
have negative chirality, which is again
determined by the Weyl nodes $\K^-_{h,e}$ at high energies.
In total, the sum of chiralities of all the Weyl nodes, including
the original band structure ones and the interaction-induced ones,
remain zero as required by the Nielsen-Ninomiya theorem
\cite{Nielsen1981a,Nielsen1981b}.

\begin{figure}[htbp]
\subfigure[]{\centering \epsfig{file=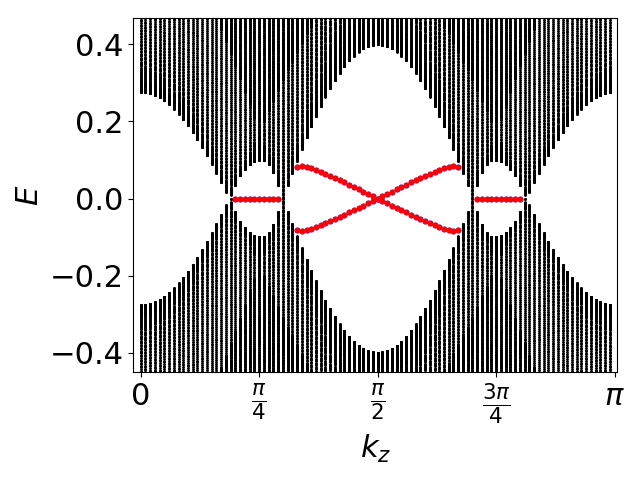,width=0.49\linewidth}}
\subfigure[]{\centering \epsfig{file=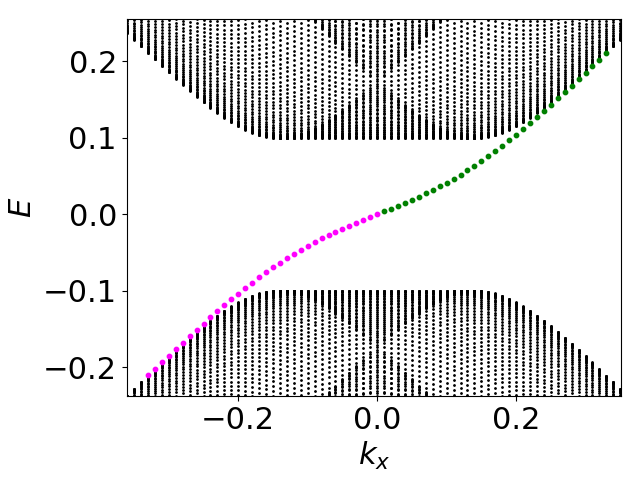,width=0.49\linewidth}}
\subfigure[]{\centering\epsfig{file=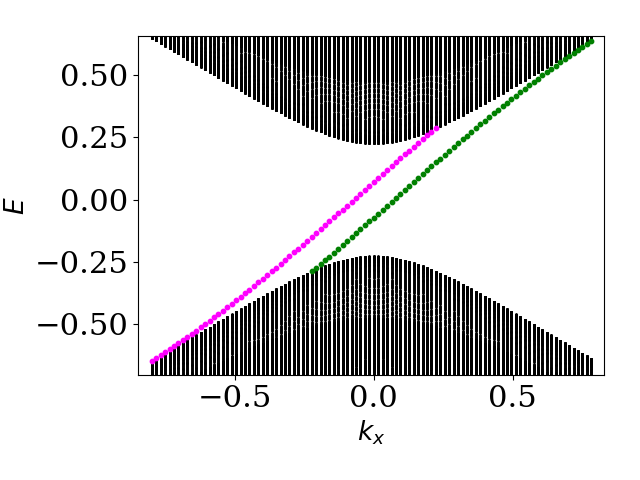,width=0.32\linewidth}}
\subfigure[]{\centering\epsfig{file=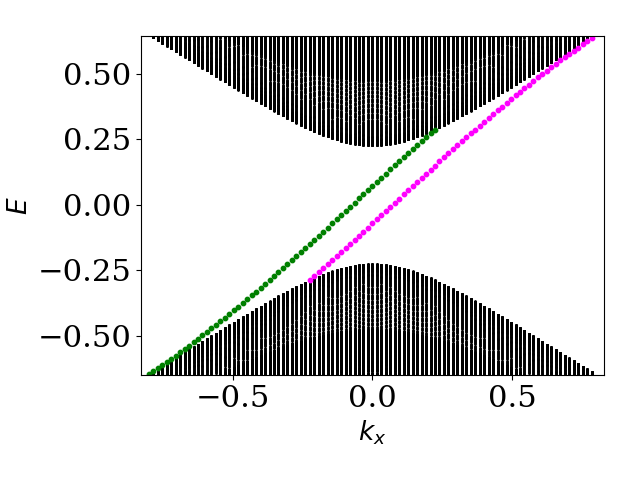,width=0.32\linewidth}}
\subfigure[]{\centering\epsfig{file=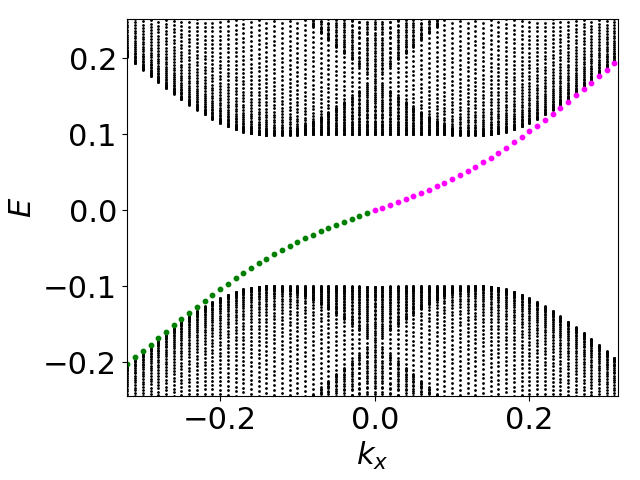,width=0.32\linewidth}}
\caption{
The bulk and surface spectra for the topological CDW ordering
with emergent Weyl nodes of Eq. \eqref{eq:cdw}.
The open boundaries are perpendicular to the $y$-axis, and only
surface states localized at the $y=0$ boundary are shown.
Parameter values are $\rho/t = 0.1$ and $V_0/t = 0.2$.
a) The dispersions along the cut at $k_x=0$ with varying $k_z$
in the reduced BZ with $0\le k_z\le \pi$.
$\K_e^-$ and $\K_h^+$ at $k_z<0$ are folded into the reduced BZ.
Surface state spectra are plotted in red.
The dispersions for varying $k_x$ are shown for constant $k_z$ cuts at
$b$) $k_z=\frac{\pi}{4}$, $c$) $k_z=\frac{3\pi}{8}$,
$d$) $k_z=\frac{5\pi}{8}$, and $e$) $k_z=\frac{3\pi}{4}$.
Green and magenta respectively indicate surface states with majority weight
in the $0\le k_z \le \pi$ and $-\pi \le k_z \le 0$ components.
}
\label{fig:topo_cdw}
\end{figure}

The emergent Weyl nodes in the monopole CDW gap function as well as novel topological surface states are demonstrated in the quasi-particle energy spectra in Fig. \ref{fig:topo_cdw}, where we take open boundary conditions along $y$-direction and periodic boundary conditions along $x$ and $z$ directions.
Because of the nesting vector $\Q=(0,0,\pi)$, the reduced Brillouin zone (BZ) with $0 \le k_z \le \pi$ is considered.
As shown in Fig.  \ref{fig:topo_cdw} ($a$), two pairs of emergent zero-energy Weyl nodes $\K^-_{s,n}$ and $\K^+_{s,n}$ appear at the $k_z$-axis near $\K_h^-$ and $\K_e^+$ in the bulk quasi-particle energy spectrum.
The surface states localized at the $y=0$ boundary are shown in color. 
Away from the Fermi surface, there are two branches of chiral surface states for $(\K^-_n)_z<k_z<(\K^+_s)_z$, due to BZ folding of the original Fermi arcs in Fig. \ref{fig:cdw_band} (b).
The number of branches of surface states changes as $k_z$ moves across
$\K^-_s$, $\K^-_n$, $\K^+_s$ and $\K^+_n$, as shown in Fig.
\ref{fig:topo_cdw} ($b$) $\sim$ ($e$).
The surface states inside the CDW gaps with $(\K^{\pm}_s)_z<k_z<(\K^{\pm}_n)_z$ are quasi-particles that are superpositions of electron states at $\k$ and $\k +\Q$ due to the particle-hole pairing near the Fermi surface. The component with majority weight changes when these surface states cross zero energy in Fig. \ref{fig:topo_cdw} ($b$) and ($e$), in contrast to the surface states shown in ($c$) and ($d$) where each branch of surface states is dominated by a single component. Thus, the CDW gaps admit novel zero-energy quasi-particles as equal superpositions of $\k$ and $\k + \Q$ states, that are a CDW analog of Majorana modes. These novel zero-energy modes are symmetric under the $\k \leftrightarrow \k + \Q$ symmetry of the Hamiltonian $H$, analogous to the charge conjugation symmetry of Majorana modes.

\begin{figure}[tbp]
\subfigure[]{\centering \epsfig{file=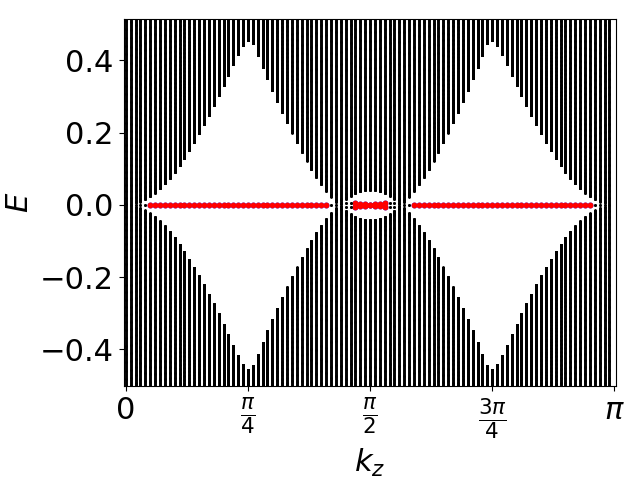,width=0.48\linewidth}}
\subfigure[]{\centering \epsfig{file=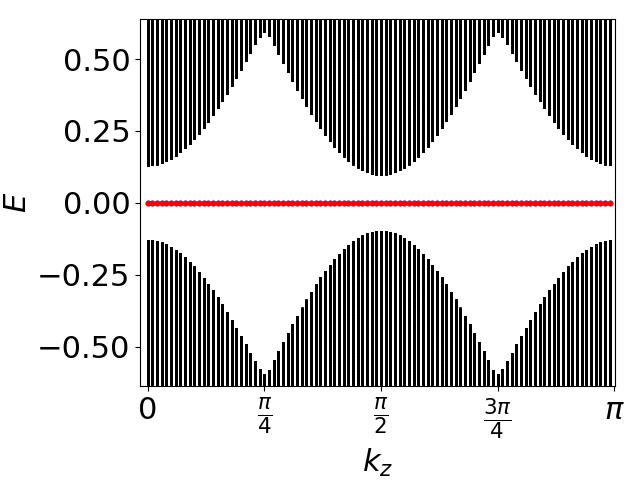,width=0.48\linewidth}}
\caption{Evolution of the low-energy Weyl nodes as the ordering
strength $\rho$ is increased with fixed $V=0.2t$.
a) At $\rho\approx0.458t$, a gap at $k_z=0$ starts to open as $\K^-_s$ merges with $\K^+_n$.
b) By $\rho=0.6t$, there is a gap at all $k_z$ after $\K^-_n$ has merged
with $\K^-_s$ at $k_z = \frac{\pi}{2}$.
}
\label{fig:bec}
\end{figure}

A natural question is whether the projection to the low-energy
helical Fermi surface remains valid as the CDW ordering strength $\rho$ becomes large.
We find that the zero-energy Weyl nodes remain robust until
$\rho$ reaches a critical value $\rho_c$.
As $\rho$ is increased, the low-energy Weyl nodes $\K^-_n$ and
$\K^+_s$ are pushed closer to each other, as are $\K^-_s$
and $\K^+_n$.
As shown in Fig. \ref{fig:bec} ($a$), $\K^-_s$ and $\K^+_n$
first merge at $k_z=0$, opening a gap, and $\K^-_n$ and $\K^+_s$ remain separated.
Further enlarging $\rho$, $\K^-_n$ and $\K^+_s$ merge next at $k_z = \frac{\pi}{2}$
after which the gap opens. After both pairs of zero-energy Weyl nodes merge, the system
becomes fully gapped around zero energy, as shown in Fig. \ref{fig:bec}($b$). The surface states in this case cross the bulk gap as $k_x$ is varied.

Increasing $\rho$ beyond the point at which the system first becomes gapped, there must eventually be an additional gap closing followed by a gap opening. To see this, consider that in the $\rho \to \infty$ limit, the spectrum must be concentrated close to $\pm \rho$, the eigenvalues of $H_\rho$. In this large $\rho$ limit, there can then be no surface states crossing the gap, and the surface states shown in Fig. \ref{fig:bec}($b$) must eventually disappear following a second transition as $\rho$ increases beyond $\rho = 0.6t$. A more detailed explanation is presented in S.M. II. This process is similar to the transition
between weak- and strong-coupling topological superconductivity,
in which the strong-coupling state is topologically trivial.

{\it Connection to experiments.} --
The general requirement for material realizations of monopole CDW states is the presence of nested Fermi surfaces enclosing Weyl points of the same chirality, which can occur in either an inversion-breaking or, as in this work, a time-reversal-breaking Weyl semimetal. In addition to the intrinsic ordering mechanism, CDW states can also be driven by an ultra-fast laser pulse, as demonstrated in a layered chalcogenide system \cite{Sun2018}. If the CDW state instead occurs between Fermi surfaces enclosing Weyl points of opposite chiralities, the gap function has zero Berry phase and can either be fully gapped or develop nodes. In such a case, the symmetry of the gap function is represented by conventional spherical harmonics or their lattice counterpart and is determined energetically by specific ordering mechanisms.

To experimentally characterize the monopole CDW, phase sensitive bulk detection, which can be provided by Resonant Inelastic X-ray Scattering (RIXS), is desired to extract the non-trivial total vorticity of phase winding in the gap function. It has been shown that, 
with appropriate choices of polarization, the RIXS intensity exhibits nodal lines in a momentum transfer q-plane \cite{Kourtis2016} when the phase winding of a single-particle Weyl Hamiltonian in the same momentum plane 
is nontrivial. This result is insensitive to the energy resolution within an energy window such that the Weyl bands around different Weyl points are well separated in momentum space. Hence, RIXS is expected to probe the emergent zero-energy Weyl nodes and their phase windings in the monopole CDW ordering. Furthermore, it can map out the non-trivial total vorticity of the gap function, which is absent in the usual gapped or even nodal CDW states. Detailed analysis will be deferred to a future publication.

{\it Discussion.} -- It would be interesting to investigate other orderings besides the monopole CDW in this system. Since most Weyl semimetals are not strongly correlated, we focus on Fermi surface instability, which can give rise to charge or spin density wave states and superconductivity in the particle-hole and particle-particle channels respectively. In this model, the superconductivity arising from zero-momentum pairing occurs between Fermi surfaces with opposite Chern numbers and is thus an example of the recently proposed monopole superconductivity \cite{Li2018}. This superconducting gap function, like the density wave gap functions, is nodal and characterized by monopole harmonics. It is possible to only have the CDW instability without superconductivity or, if their energy scales are comparable, to have both instabilities coexist. Analogous to the coexistence of the CDW and superconductivity seen in the transition-metal dichalcogenide NbSe$_2$ \cite{Takita1985, Yokoya2001, Xi2015, Lian2018, Zheng2019}, competition between monopole harmonic orders is possible, and this interesting problem will be studied in future work.


{\it Conclusions.} -- We have studied the inter-Fermi-surface particle-hole
pairing between well-nested Fermi surfaces enclosing Weyl nodes of the same chirality.
The CDW ordering in this case gives rise to gap functions possessing topologically protected nodes independent of the details of ordering interactions.
These nodes manifest in the bulk single-particle energy spectra as zero-energy Weyl nodes, which allow the topology of the surface Fermi arcs to change.
The nodal structure corresponds to a new, novel type of topological CDW state whose order parameter carries monopole symmetries.
As the ordering strength increases to the strong coupling regime, the zero energy Weyl nodes merge and disappear, and the system becomes fully gapped.

{\it Acknowledgments.} --
We thank Yuxuan Wang, Peter Abbamonte, Collin Broholm, Stefanos Kourtis and Tyrel McQueen for helpful discussion.
This work is supported by the NSF Career Grant No. 1848349.
Y. L. also acknowledges the support from the Alfred P. Sloan Research Fellowships.
E.B. and C.S. contributed equally to this work.

%

\balancecolsandclearpage
\section{Supplemental Materials}
\twocolumngrid
  \pagenumbering{arabic}
  \renewcommand{\thepage}{S-\arabic{page}}
\setcounter{equation}{0}
\setcounter{figure}{0}
\setcounter{table}{0}
\setcounter{page}{1}
\makeatletter
\renewcommand{\theequation}{S\arabic{equation}}
\renewcommand{\thefigure}{S\arabic{figure}}

\section{I. Lattice Model of the Hamiltonian}
The Hamiltonian given by Eqs.  (\ref{eq:Hamiltonian})
, (\ref{eq:bloch}), and
(\ref{eq:v})
in the main text can be constructed from a real space bipartite lattice model. After a Fourier transformation, the band Hamiltonian becomes
\begin{align}
H=\sum_{\substack{\r, \\ a,b = A,B}} &\Bigg\{\left[c^{\dagger}_a(\r+\ehat_x)\left(-\frac{t_x}{2i}\tau_x-\frac{t_z}{2}\tau_z \right)_{ab}c_b(\r)+h.c.\right] \nn \\
+&\left[c^{\dagger}_a(\r+\ehat_y)\left(-\frac{t_y}{2i}\tau_y-\frac{t_z}{2}\tau_z \right)_{ab}c_b(\r)+h.c.\right] \nn \\
+&\left[c^{\dagger}_a(\r+\ehat_z)\left(\frac{V_0}{2}I \right)_{ab}c_b(\r)+h.c.\right] \nn \\
+&\left[c^{\dagger}_a(\r+ 2 \ehat_z)\left(\frac{t_z}{4}\tau_z\right)_{ab}c_b(\r)+h.c.\right] \nn \\
+&\left[c^{\dagger}_a(\r)\left(t_x\left(5/2-\gamma \right)\tau_z \right)_{ab}c_b(\r)\right]\Bigg\}, \nn
\end{align}
where $c^\dagger_a(\r)$ and $c_a(\r)$ are respectively the electron creation and annihilation operators on site $\r$ with sublattice $a=A,B$. $\ehat_i$ ($i=x,y,z$) are the unit lattice vectors pointing in the $i$ direction and $\tau_i$ are Pauli matrices in the sublattice basis. The first three terms describe nearest neighbor hopping in the $x$-, $y$-, and $z$- directions with sublattice dependent hopping amplitudes $t_i$ and $V_0$. The fourth is a next nearest neighbor hopping term in the $z$-direction. The last term describes a sublattice modulated on-site potential with parameter $\gamma$, which controls the position of the Weyl points along the $k_z$ axis.

\section{II. Strong-Coupling CDW}

As discussed in the main text, there can be no surface states crossing the gap once $\rho$ is sufficiently large. This can be seen explicitly by considering the Hamiltonian in a BdG-like form,
\begin{equation}
\begin{aligned}
h_{BdG}(\k) &= V_0 \cos k_z \sigma_z \otimes I + \sin k_x I \otimes \tau_x + \sin k_y I \otimes \tau_y \\
&+ \left(\frac{3}{2} - \cos k_x - \cos k_y + \cos^2k_z\right)I \otimes \tau_z \\
&+ \rho \sigma_x \otimes \tau_z.
\end{aligned}
\end{equation}
As in the main text, $t_{x,y,z} = t$ and $V_0$ and $\rho$ are expressed in units of $t$. The $\tau$ matrices are pseudospin Pauli matrices corresponding to the $A, B$ sublattice degrees of freedom, while the $\sigma$ Pauli matrices correspond to the $\k$ and $\k + \Q$ parts of the BdG basis $[c_A(\k), c_B(\k), c_A(\k + \Q), c_B(\k + \Q)]^T$. By explicitly diagonalizing the BdG Hamiltonian, it can be shown that the gap can close only when $k_x, k_y = 0$ or $\pi$, where the energy spectrum can be written
\begin{equation}
E(\k)_{\pm, \pm} = \pm(\g_\k + \cos^2k_z) \pm\sqrt{\rho^2 + V_0^2 \cos^2k_z}
\end{equation}
with
\begin{equation}
\g_\k = \begin{cases}\;\;\;\frac{7}{2} & (k_x, k_y) = (\pi, \pi)\\
\;\;\;\frac{3}{2} & (k_x, k_y) = (0, \pi), (\pi, 0) \\
 -\frac{1}{2} & (k_x, k_y) = (0, 0).
\end{cases}
\end{equation}
Critical values of $\rho$ for which the gap closes somewhere in the BZ are given by
\begin{equation}
\rho_c(\k) = \sqrt{(\g_\k + \cos^2k_z)^2 - V_0^2 \cos^2 k_z}.
\end{equation}
For open boundary conditions in the $y$ direction and periodic boundary conditions in $k_x$ and $k_z$, a gap closing will be seen in the spectrum at $(k_x, k_z)$ if there is a positive critical $\rho_c(\k)$ at $(k_x, k_z)$ for any $k_y$.

\begin{figure}[tbp]
\subfigure{\centering \epsfig{file=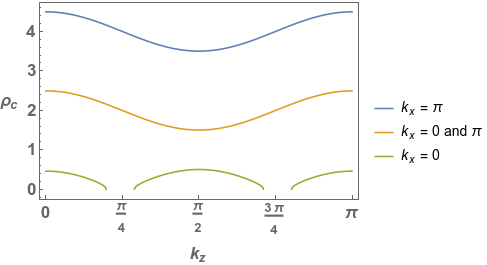,width=0.98\linewidth}}
\caption{Curves of the critical coupling $\rho_c$ at which the gap closes as a function of $k_z$ with $V_0 = 0.2$. The top curve describes the gap closing at $k_x = \pi$, while the bottom curve describes $k_x = 0$. Along the middle curve, there are simultaneous gap closings at both $k_x = 0$ and $k_x = \pi$. The gap does not close for positive $\rho$ at any other values of $k_x$.}
\label{fig:rhoc}
\end{figure}

The critical values $\rho_c$ as a function of $k_z$ are shown in Fig. \ref{fig:rhoc} for $V_0 = 0.2$. There are only three critical curves, as there can only be gap closings at $k_x = 0$ or $\pi$, giving only three possible values for $\g_\k$. The middle curve contains information about both $k_x = 0$ and $k_x = \pi$, though these transitions are qualitatively different. The first transition mentioned in the main text corresponds to the bottom curve as $\rho$ is increased from $0$ to $0.5$. For $\rho$ between the bottom and middle critical curves, there is an edge state extending across all $k_z$ in the $k_x = 0$ cut as shown in Fig. \ref{fig:bec}.

Once $\rho$ begins to cross the middle curve, at $\rho_c = 1.5$, the gap closes at $k_z = \pi/2$ for both $k_x = 0$ and $k_x = \pi$. Increasing $\rho$ through the middle curve, the surface states at $k_x = 0$ begin to disappear as a gap opens from $k_z = \pi/2$ outwards until the gap nodes are pushed to $k_z = \pi$ for $\rho_c \approx 2.49$. Increasing $\rho$ further leaves the system gapped at all points along $k_x = 0$, and no surface states will cross the gap for larger $\rho$. At $k_x = \pi$ instead, crossing the middle curve introduces surface states as the gap opens from $k_z = \pi/2$ outwards. These surface states disappear after the final transition as $\rho$ is increased past the top curve, and a fully gapped system remains with no surface states crossing the gap once $\rho$ exceeds $\rho_c \approx 4.5$.

\begin{figure}[htbp]
\subfigure[]{\centering \epsfig{file=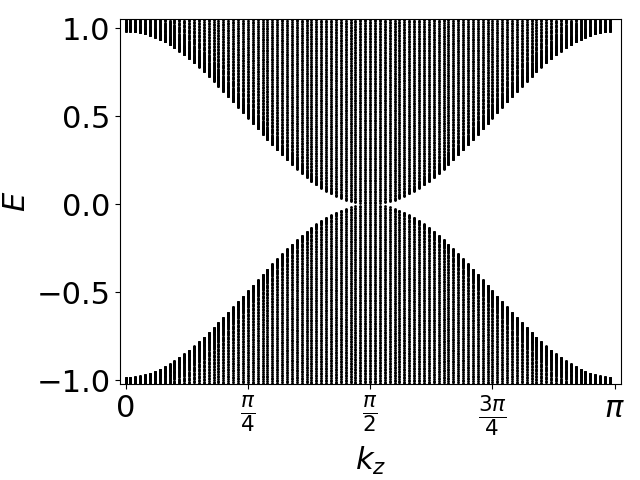,width=0.48\linewidth}}
\subfigure[]{\centering \epsfig{file=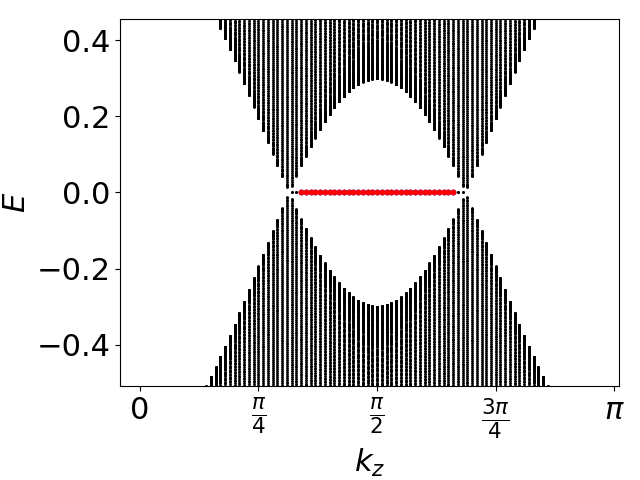,width=0.48\linewidth}}
\subfigure[]{\centering \epsfig{file=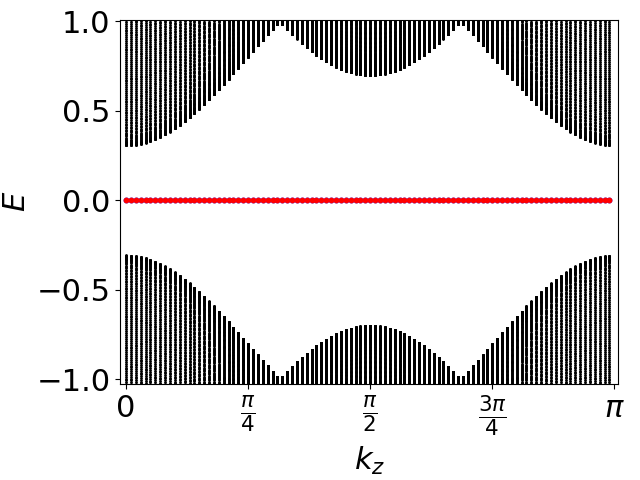,width=0.48\linewidth}}
\subfigure[]{\centering \epsfig{file=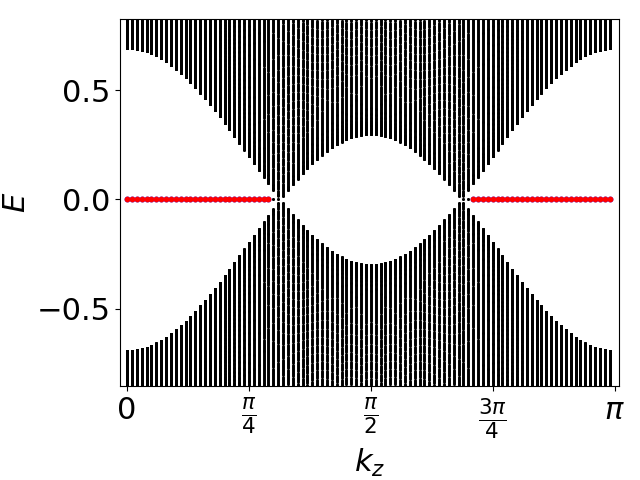,width=0.48\linewidth}}
\caption{Gap closings as $\rho$ is varied at $k_x = \pi$. a) At $\rho = 1.5$, the gap closes at $\k_z = \pi/2$. b) $\rho = 1.8$. As $\rho$ is increased past the initial gap closing, a gap with surface states opens and the gap nodes are pushed outwards. c) $\rho = 2.8$. Once the gap nodes reach $\k_z = \pi$, the gap opens leaving a surface state. d) $\rho = 3.8$. As $\rho$ increases further, the gap closes at $k_z = \pi/2$ again and reopens, pushing the gap nodes towards $k_z = \pi$ and eliminating the surface states.
}
\label{fig:kxpi}
\end{figure}

The gap closings at $k_x = \pi$ are shown for $\rho$ crossing the middle and top critical curves in Fig. \ref{fig:kxpi}. The transition that occurs at $k_x = \pi$ as $\rho$ crosses the top curve is qualitatively similar to the transition at $k_x = 0$ as $\rho$ crosses the middle curve.

\end{document}